\def\be{\begin{equation}}
\def\ee{\end{equation}}

\documentclass[prl,twocolumn,superscriptaddress,showpacs]{revtex4}
\usepackage{graphicx}
\usepackage{amssymb}
\usepackage{mathrsfs}
\usepackage{amsmath}

\begin{document}

\title{Cluster Monte Carlo and numerical mean field analysis for the
water liquid--liquid phase transition}

\author{Marco G. Mazza}
\affiliation{Center for Polymer Studies and Department of Physics,
  Boston University, Boston, Massachusetts 02215, USA}
\author{Kevin Stokely}
\affiliation{Center for Polymer Studies and Department of Physics,
  Boston University, Boston, Massachusetts 02215, USA}
\author{Elena Strekalova}
\affiliation{Center for Polymer Studies and Department of Physics,
  Boston University, Boston, Massachusetts 02215, USA}
\author{H. Eugene Stanley} 
\affiliation{Center for Polymer Studies and Department of Physics,
  Boston University, Boston, Massachusetts 02215, USA}
\author{Giancarlo Franzese}
\affiliation{Departament de Fisica Fonamental, 
Universitat de Barcelona, Diagonal 647, 08028 Barcelona, Spain}

\begin{abstract}
  By the Wolff's cluster Monte Carlo simulations and numerical
  minimization within a mean field approach, we study the low
  temperature phase diagram of water, adopting a cell model that
  reproduces the known properties of water in its fluid phases.  Both
  methods allows us to study the water thermodynamic behavior at
  temperatures where other numerical approaches --both Monte Carlo and
  molecular dynamics-- are seriously hampered by the large increase of
  the correlation times.  The cluster algorithm also allows us to
  emphasize that the liquid--liquid phase transition corresponds to
  the percolation transition of tetrahedrally ordered water molecules.

\end{abstract}
\maketitle

\section{Introduction}

Water is possibly the most important liquid for life \cite{sitges}
and, at the same time, is a very peculiar liquid
\cite{Debenedetti-JPCM03}. In the stable liquid regime its 
thermodynamic response functions behave qualitatively differently than
a typical liquid. The isothermal compressibility $K_T$, for example,
has a minimum as a function of temperature at $T=46~^\circ$C, while for
a typical liquid $K_T$ monotonically decreases upon cooling. Water's
anomalies become even more pronounced as the system is cooled below
the melting point and enters the metastable supercooled regime
\cite{DebenedettiStanley}.

Different hypothesis have been proposed to rationalize the anomalies of
water \cite{Franzese-JPCM08}. All these interpretations, but one,
predict the existence of a liquid--liquid phase transition in the
supercooled state, consistent with the experiments to date
\cite{Franzese-JPCM08} and supported by different models
\cite{Debenedetti-JPCM03}. 

To discriminate among the different interpretations, many experiments
have been performed \cite{angell2008}. However, the freezing in the
temperature-range of interest can be avoided only for water in
confined geometries or on the surface of macromolecules
\cite{Franzese-JPCM08,Stanley_etal}.  Since experiments in the
supercooled region are difficult to perform, numerical simulations
have played an important role in recent years to help interpret
the data. However, also the simulations at very low temperature $T$
are hampered by the glassy dynamics of the empirical models of
water~\cite{slow,kfsPRL2008}.  For these reasons is important to
implement more efficient numerical simulations for simple models, able
to capture the fundamental physics of water but also less
computationally expensive.  Here we introduce the implementation of a
Wolff's cluster algorithm~\cite{wolff} for the Monte Carlo (MC)
simulations of a cell model for water~\cite{fs}.  The model is able to
reproduce all the different scenarios proposed to interpret the
behavior of water~\cite{kevin} and has been analyzed (i) with mean
field (MF) \cite{fs,Franzese-JPCM07,kumar-JPCM08}, (ii) with
Metropolis MC simulations \cite{fmsPRE03,kfsPRL2008} and (iii) with
Wang-Landau MC density of state algorithm \cite{marques-PRE07}. Recent
Metropolis MC simulations \cite{kfsPRL2008} have shown that very large
times are needed to equilibrate the system as $T\rightarrow 0$, as a
consequence of the onset of the glassy dynamics.  The implementation
of the Wolff's clusters MC dynamics, presented here, allows us to (i)
drastically reduce the equilibration times of the model at very low
$T$ and (ii) give a geometrical characterization of the regions of
correlated water molecules (clusters) at low $T$ and show that the
liquid--liquid phase transition can be interpreted as a percolation
transition of the tetrahedrally ordered clusters.

\begin{figure}
\includegraphics[scale=0.55]{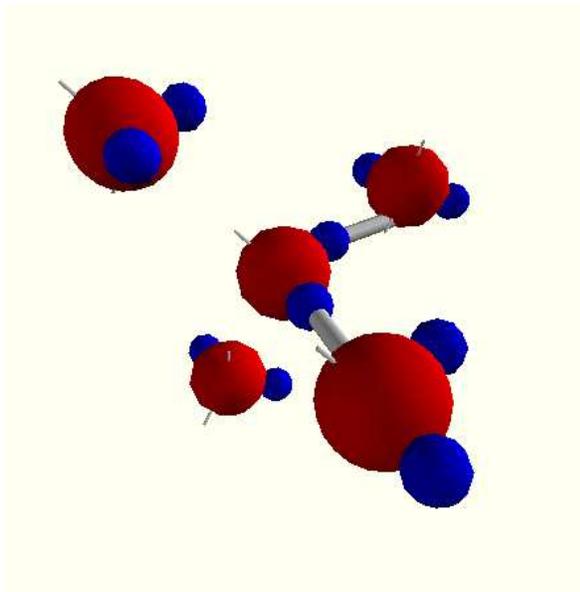}
\caption{A pictorial representation of five
water molecules in 3$d$. Two hydrogen bonds (grey links) connect the
hydrogens (in blue) of the central molecule with the lone electrons
(small gray lines) 
of two nearest neighbor (n.n.) molecules. 
A bond index (arm) with $q=6$ possible values is associated to each
hydrogen and lone electron, giving rise to $q^4$
possible orientational states for each molecule. 
A hydrogen bond can be formed only if
the two facing arms of the n.n. molecules are in the same state.
Arms on the same molecule interact among themselves to mimic the O-O-O
interaction that drives the molecules toward a tetrahedral 
local structure.}
\label{picture}
\end{figure}

\section{The model}

The system consists of $N$ particles distributed within a volume $V$
in $d$ dimensions. The volume is divided into $N$ cells of volume
$v_i$ with $i\in[1,N]$. For sake of simplicity, these cells are chosen
of the same size, $v_i=V/N$, but the generalization to the case in
which the volume can change without changes in the topology of the
nearest--neighbor (n.n.) is straightforward. By definition, $v_i\geq
v_0$, where $v_0$ is the molecule hard-core volume.
Each cell has a variable
$n_i=0$ for a gas-like or $n_i=1$ for a liquid-like cell.  
We partition the total volume in a way such that each cell has at
least four n.n. cells, e.g. as in a cubic lattice in 3$d$ or a square
lattice in 2$d$. Periodic boundary conditions are used to limit
finite--size effects.

The system is described by the Hamiltonian \cite{fs}
\begin{multline}
  \mathscr{H} =   - \epsilon \sum_{\langle i,j \rangle} n_i n_j 
-J \sum_{\langle ij \rangle} n_i n_j
  \delta_{\sigma_{ij},\sigma_{ji}} +\\
  -J_\sigma \sum_i n_i \sum_{(k,l)_i} \delta_{\sigma_{ik},\sigma_{il}},
\label{ham}
\end{multline}
\noindent
where $\epsilon>0$ is the strength of the van der Waals attraction,
$J>0$ accounts for the hydrogen bond energy, with four (Potts)
variables $\sigma_{ij}=1,\ldots,q$ representing 
bond indices of molecule $i$ with respect to the four n.n. molecules
$j$, $\delta_{a,b}=1$ if $a=b$ and $\delta_{a,b}=0$ otherwise, and
$\langle i,j\rangle$ denotes that $i$ and $j$ are n.n. 
The
model does not assume a privileged state for bond formation. Any 
time two facing bond indices (arms) are in the same (Potts) state, a
bond is formed. The third term
represents an intramolecular (IM) interaction accounting for the O--O--O
correlation \cite{Ricci-Chaplin}, locally driving the molecules toward a
tetrahedral configuration.
When the bond indices of a molecule are in the same 
state, the energy
is decreased by an amount $J_\sigma\geqslant0$ and we associate this
local ordered configuration to a local tetrahedral arrangement \cite{note}. 
The notation $(k,l)_i$
indicates one of the six different pairs of the four bond indices of
molecule $i$ (Fig.\ref{picture}).  

Experiments show that the formation of a hydrogen bond leads to a
local volume expansion~\cite{Debenedetti-JPCM03}. Thus in our system
the total volume is 
\begin{equation}
\label{vol}
V = N v_0 + N_{HB} v_{HB},
\end{equation}
where 
\be 
N_{HB}\equiv\sum_{<i,j>} n_i n_j  \delta_{\sigma_{ij},\sigma_{ji}} 
\ee
is the total number of hydrogen bonds, and $v_{HB}$ is the constant
specific volume increase due to the hydrogen bond formation.

\section{Mean--field analysis}

In the mean--field (MF) analysis the macrostate of the
system in equilibrium at constant pressure $P$ and temperature $T$
($NPT$ ensemble) may be determined by a minimization of the Gibbs
free energy per molecule, $g \equiv (\langle \mathscr{H} \rangle + PV -
TS)/N_w$, where 
\begin{equation}
\label{num}
N_w = \sum_i n_i 
\end{equation}
is the total number of liquid-like cells, and 
$S=S_n+S_\sigma$ is the sum of the entropy $S_n$
over the variables $n_i$ and the entropy $S_\sigma$ over the variables
$\sigma_{ij}$. 

A MF approach consists of writing $g$ explicitly using the approximations
\begin{align}
\sum_{<ij>} n_i n_j &\longrightarrow 2 N n^2 \\
\sum_{<ij>} n_i n_j \delta_{\sigma_{ij}, \sigma_{ji}} &\longrightarrow 2 N n^2 p_\sigma \\
\sum_{i} n_i \sum_{(k,l)_i} \delta_{\sigma_{ik}, \sigma_{il}}
&\longrightarrow 6 N n p_\sigma
\end{align}
\noindent
where $n=N_w/N$ is the average of $n_i$,
and $p_\sigma$ is
the probability that two adjacent bond indices $\sigma_{ij}$
are in the
appropriate state to form a hydrogen bond.  

Therefore, in this approximation we can write
\begin{align}
V & = N v_0 + 2 N n^2 p_\sigma v_{HB},\\
\langle \mathscr{H} \rangle & = -2\left[\epsilon n+\left(J n+ 3
J_\sigma\right)p_\sigma \right] n N .
\end{align}

The probability
$p_\sigma$, properly defined as the thermodynamic average over the
whole system, is approximated as 
the average over two neighboring molecules, under the effect of the
mean-field $h$ of the surrounding molecules
\begin{equation}
\label{psig1}
p_\sigma = \left< \delta_{\sigma_{ij},\sigma_{ji}} \right>_h .
\end{equation}

The ground state of the system consists of all $N$ variables $n_i=1$,
and all $\sigma_{ij}$ in the same state.  At low temperatures,
the symmetry will remain broken, with the majority of the
$\sigma_{ij}$ in the preferred state. We associate this
preferred state to the tetrahedral order of the molecules and define 
$m_\sigma$ as the density of the bond indices
in the tetrahedral state, with $0 \leq m_\sigma \leq 1$.
Therefore, the number density $n_\sigma$ of bond indices 
$\sigma_{ij}$ is in the tetrahedral state is 
\begin{equation}
n_\sigma={{1 + (q-1)m_\sigma} \over q}.
\end{equation}

Since an appropriate form for $h$ is \cite{fs}
\begin{equation}
h = 3 J_\sigma n_\sigma,
\end{equation}
we obtain that ${{3J_\sigma} \over q} \leq h \leq 3J_\sigma$. 
 
The MF expressions for the entropies $S_n$ of the $N$ variables
$n_i$, and $S_\sigma$ of the $4Nn$ variables $\sigma_{ij}$, are then
\cite{Franzese-JPCM07} 
\be
\shoveleft{S_n = -k_BN(n\log(n) + (1-n)\log(1-n))}
\ee
\begin{multline}
S_\sigma = -k_B4Nn[n_\sigma\log(n_\sigma) + \\
(1-n_\sigma)\log(1-n_\sigma) + \log(q-1)] ,
\end{multline}
where $k_B$ is the Boltzmann constant.
 
Equating 
\begin{equation}
\label{psig2}
p_\sigma \equiv  n_\sigma^2 + {{(1-n_\sigma)^2} \over {q-1}} ,
\end{equation}
with the approximate expression in 
Eq. (\ref{psig1}), allows for solution
of $n_\sigma$, and hence $g$, in terms of the order parameter
$m_\sigma$ and $n$.

\begin{figure}
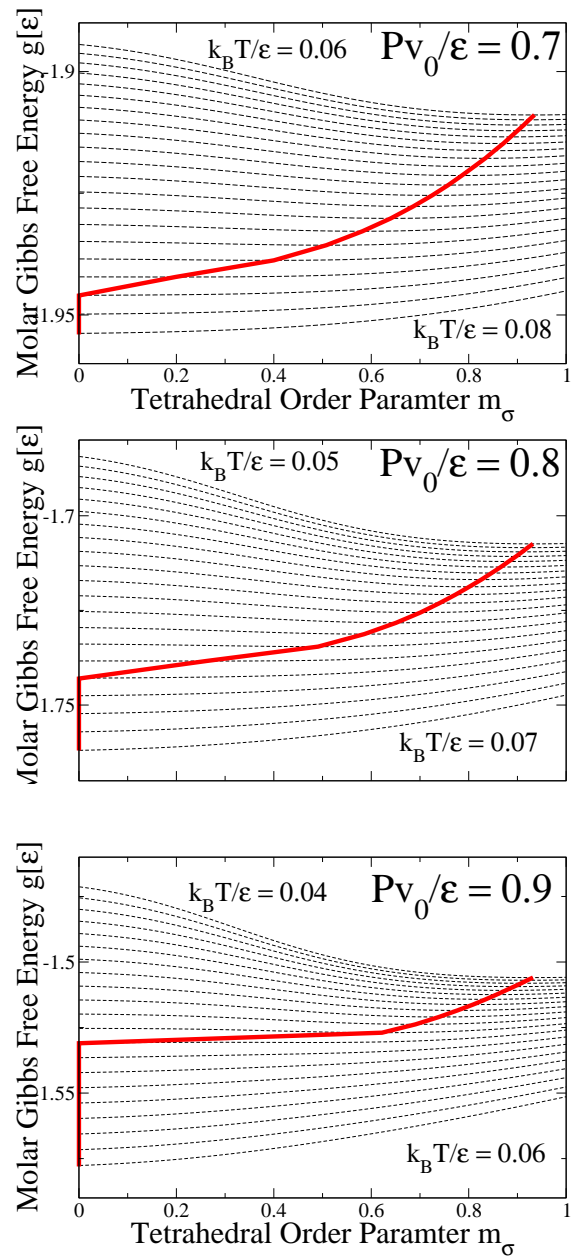

\includegraphics[scale=0.3]{mssf-CPC-fig2a.eps}
\includegraphics[scale=0.3]{mssf-CPC-fig2b.eps}
\includegraphics[scale=0.3]{mssf-CPC-fig2c.eps}
\caption{Numerical minimization of the molar 
Gibbs free energy $g$ in the mean field approach. The model's
parameters are $J/\epsilon=0.5$,  
$J_\sigma/\epsilon=0.05$, $v_{HB}/v_0=0.5$ and $q=6$.
In each panel we present $g$ (dashed lines) calculated at constant $P$
and different values of $T$. The thick line crossing the dashed lines
connects the minima $m_\sigma^{(eq)}$ of $g$ at different $T$.
Upper panel: $Pv_0/\epsilon=0.7$, for $T$ going from
$k_BT/\epsilon=0.06$ (top) to $k_BT/\epsilon=0.08$ (bottom).
Middle panel: $Pv_0/\epsilon=0.8$, for $T$ going from
$k_BT/\epsilon=0.05$ (top) to $k_BT/\epsilon=0.07$ (bottom).
Lower panel: $Pv_0/\epsilon=0.9$, for $T$ going from
$k_BT/\epsilon=0.04$ (top) to $k_BT/\epsilon=0.06$ (bottom).
In each panel dashed lines are separated by $k_B \delta
T/\epsilon=0.001$. In all the panels
  $m_\sigma^{(eq)}$ increases when $T$ decreases, being 0 (marking
  the absence of tetrahedral order) at the
  higher temperatures and $\simeq 0.9$ (high tetrahedral order) at the
  lowest temperature. By changing $T$, $m_\sigma^{(eq)}$ changes in
  a continuous way for $Pv_0/\epsilon=0.7$ and $0.8$, but
  discontinuous for $Pv_0/\epsilon=0.9$ and higher $P$.
}
\label{mf}
\end{figure}

By minimizing numerically the MF expression of $g$
with respect to $n$ and $m_\sigma$, we find the 
equilibrium values $n^{(eq)}$ and $m_\sigma^{(eq)}$ and, with 
Eqs. (\ref{num}) and (\ref{vol}), we calculate the density
$\rho$ at any $(T,P)$ and the full equation of state. An example of
minimization of $g$ is presented in Fig.~\ref{mf} where, for the
model's parameters $J/\epsilon=0.5$,  
$J_\sigma/\epsilon=0.05$, $v_{HB}/v_0=0.5$, $q=6$, a discontinuity in
$m_\sigma^{(eq)}$ is observed for $Pv_0/\epsilon>0.8$. As discussed in
Ref.s~\cite{fs,fmsPRE03} this discontinuity corresponds to a first
order phase transition between two liquid phases with different degree
of tetrahedral order and, as a consequence, different density. The
higher $P$ at which the change in $m_\sigma^{(eq)}$ is continuous,
corresponds to the pressure of a liquid--liquid critical point
(LLCP). The occurrence of the LLCP is consistent with one of the
possible interpretations of the anomalies of water, as discussed in
Ref.~\cite{Franzese-JPCM07}. However, for different choices of
parameters, the model reproduces also the other proposed scenarios
\cite{kevin}. 

\section{The simulation with the Wolff's clusters Monte Carlo algorithm}

To perform MC simulations in the $NPT$ ensemble, we consider a
modified version of the model in which we allow for continuous volume
fluctuations. To this goal, (i) we assume that the system is
homogeneous with all the variables $n_i$ set to 1 and all the cells
with volume $v=V/N$; (ii) we consider that 
$V\equiv V_{MC}+N_{HB}v_{HB}$, where
$V_{MC}\geqslant N v_0$ is a dynamical variable allowed to fluctuate
in the simulations; (iii) we replace the first (van der Waals) term of
the Hamiltonian in Eq. (\ref{ham}) with a Lennard-Jones potential with
attractive energy $\epsilon>J$ and truncated at the hard-core distance
\be
U_W(r)\equiv\begin{cases}
\infty & \text{if $r\leqslant r_0$,}\\
\epsilon\left[\left(\frac{r_0}{r}\right)^{12}-
      \left(\frac{r_0}{r}\right)^6\right] & \text{if $r>r_0$.}\\
\end{cases}
\label{LJ}
\ee
where $r_0\equiv(v_0)^{1/d}$; the distance between two n.n. molecules
is $(V/N)^{1/d}$, and the distance $r$ between two generic molecules 
is the Cartesian distance between the center of the cells in which
they are included.

The simplification (i) could be removed, by allowing the cells to
assume different volumes $v_i$ and keeping fixed the number of
possible n.n. cells. However, the results of the model under the
simplification (i) compares well with experiments
\cite{Franzese-JPCM07}.  Furthermore, the simplification (i) allows to
drastically reduce the computational cost of the evaluation of the
$U_W(r)$ term from $N(N-1)$ to $N-1$ operations.  The changes
(i)--(iii) modify the model used for the mean field analysis and allow
off-lattice MC simulations for a cell model in which the topology of
the molecules (i.e. the number of n.n.) is preserved.  The comparison
of the mean field results with the MC simulations show that these
changes do not modify the physics of the system.

We perform MC simulations with $N=2500$ and $N=10000$ molecules, 
each with four n.n. molecules, at constant $P$ and $T$, in 2d, and
with the same parameters used for the mean field analysis.
To each molecules we associate a cell on a square lattice.
The Wolff's algorithm is based on the definition of a cluster of variables 
chosen in such a way to be thermodynamically correlated
\cite{clusters,coniglio}.
To define the Wolff's cluster, a bond index (arm) of a molecule is randomly
selected; this is 
the initial element of a stack. The cluster 
is grown by first checking the remaining arms of the 
same initial molecule: if they are in the same Potts state, then they are
added to the stack with probability $p_{\rm same}\equiv
\min\left[1,1-\exp(-\beta J_\sigma)\right]$ \cite{wolff}, where
$\beta\equiv(k_BT)^{-1}$.
This choice for the probability $p_{\rm same}$
depends on the interaction $J_\sigma$ between two
arms on the same molecule and guarantees that the connected arms are
thermodynamically correlated \cite{coniglio}.
Next,
the arm of a new molecule, facing the initially chosen arm, is
considered. 
To guarantee that connected facing arms correspond to
thermodynamically correlated variables, is necessary \cite{clusters} to link
them with the probability 
$p_{\rm facing}\equiv \min\left[1,1-\exp(-\beta J')\right]$
where $J'\equiv J-P v_{HB}$ is the $P$--dependent effective coupling 
between two facing arms as results from the enthalpy 
$\mathscr{H}  + PV$ of the system.
It is important to note that $J'$ can be
positive or negative depending on $P$. If $J'>0$ and the two facing
arms are in the same state, then the new arm is added to the
stack with probability $p_{\rm facing}$; if $J'<0$ and the two facing
arms are in different 
states, then the new arm is added with probability
$p_{\rm facing}$ \cite{note2}. Only after every possible direction of
growth for the 
cluster has been considered the values of the arms are
changed in a stochastic way; again we need to consider two cases: (i)
if $J'>0$, all arms are 
set to the same new value 
\be
\sigma^{\rm new}=\left( \sigma^{\rm old}+\phi \right) \mod q
\ee
where $\phi$ is a random integer between 1 and $q$;
(ii) if $J'<0$, the state of every single arm is changed (rotated) by
the same random constant $\phi \in [1,\dots q]$
\be
\sigma_i^{\rm new}=\left( \sigma_i^{\rm old}+\phi\right) \mod q .
\ee

In order to implement a constant $P$ ensemble we 
let the volume fluctuate. A small increment
$\Delta r/r_0=0.01$ is chosen with uniform random probability and added to the
current radius of a cell. The change in volume $\Delta V\equiv
V^{\rm new}-V^{\rm old}$ and van der Waals energy $\Delta E_W$ is computed and
the move is accepted with probability
$\min\left(1,\exp\left[-\beta\left(\Delta E_W+P\Delta V-T \Delta
      S\right)\right]\right)$, where $\Delta S\equiv
-Nk_B\ln(V^{new}/V^{old})$ is the entropic contribution.

\section{Monte Carlo correlation times}

The cluster MC algorithm described in the previous section
turns out to be 
very efficient at low $T$, allowing to study the thermodynamics of
deeply supercooled water with quite intriguing results \cite{mazza}.
To estimate the efficiency of the cluster MC dynamics with respect to
the standard Metropolis MC dynamics,  
we evaluate in both dynamics, and compare, the autocorrelation
function of the average magnetization per site
$M_i\equiv\frac{1}{4}\sum_j\sigma_{ij}$, where the sum is over the
four bonding arms of molecule $i$.

\be
C_M(t)\equiv\frac{1}{N}\sum_i
\frac{\langle M_i(t_0+t)M_i(t_0)\rangle-\langle M_i \rangle^2} 
{\langle M_i^2 \rangle-\langle M_i \rangle^2} .
\ee

For sake of simplicity, we define the MC dynamics autocorrelation
time $\tau$ as the time, measured in MC steps, when $C_M(\tau)=1/e$. 
Here we define a MC step as $4N$ updates of the bond indices followed
by a volume update, i.e. as $4N+1$ steps of the algorithm.

\begin{figure}
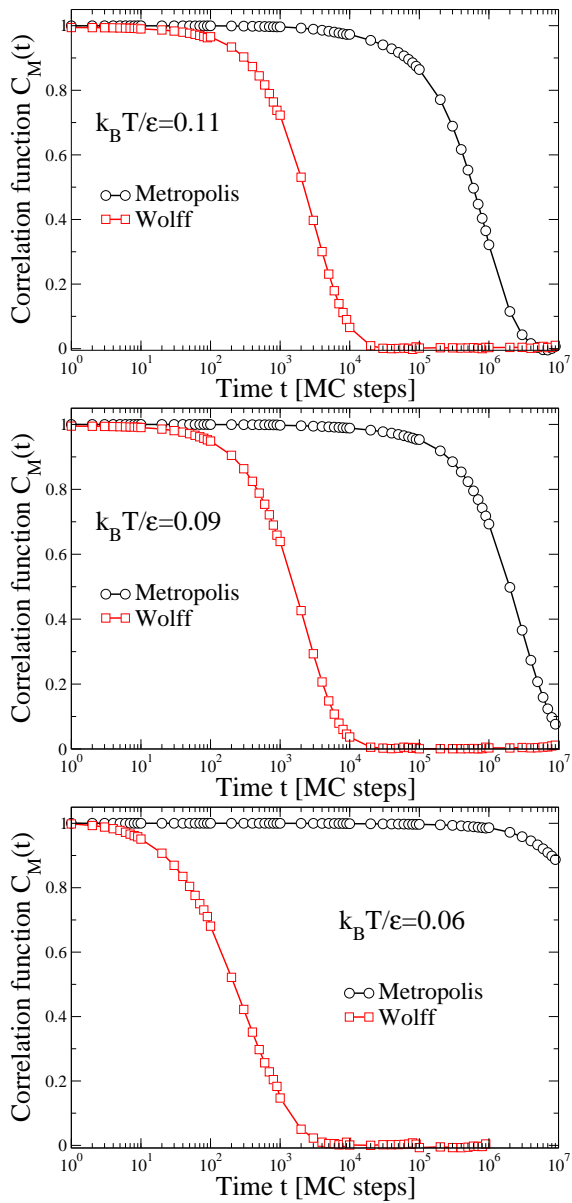

\includegraphics[scale=0.3]{T0.11-P0.6-L50.eps}
\includegraphics[scale=0.3]{T0.09-P0.6-L50.eps}
\includegraphics[scale=0.3]{T0.06-P0.6-L50.eps}
\caption{Comparison of the autocorrelation function $C_M(t)$ for the
  Metropolis (circles) and Wolff (squares) implementation of the
  present model. We show the temperatures $k_BT/\epsilon=0.11$ (top
  panel), $k_BT/\epsilon=0.09$ (middle panel), $k_BT/\epsilon=0.06$
  (bottom panel), along the isobar $Pv_0/\epsilon=0.6$ close to the
  LLCP for $N=50\times 50$.\label{corrfu}}
\end{figure}

In Fig.~\ref{corrfu} we show a comparison of $C_M(t)$ for the
Metropolis and Wolff 
algorithm implementations of this model for a system with $N=50\times
50$, at three temperatures along an
isobar below
the LLCP, and approaching the line of the maximum, but finite,
correlation length, also known as Widom line $T_W(P)$~\cite{Franzese-JPCM07}.
In the top panel, at $T\gg T_W(P)$ ($k_BT/\epsilon=0.11$,
$Pv_0/\epsilon=0.6$), 
we find a correlation time for the Wolff's cluster MC
dynamics $\tau_{\rm W}\approx3\times10^3$, and for the Metropolis
dynamics $\tau_{\rm M}\approx10^6$. In the middle panel, at $T>T_W(P)$ 
($k_BT/\epsilon=0.09$, $Pv_0/\epsilon=0.6$) the difference between the
two correlation times is larger: $\tau_{\rm W}\approx2.5\times10^3$,
$\tau_{\rm M}\approx3\times10^6$. The bottom panel, at $T\simeq T_W(P)$ 
($k_BT/\epsilon=0.06$, $Pv_0/\epsilon=0.6$) shows $\tau_{\rm
  W}\approx3.7\times10^2$, while 
$\tau_{\rm M}$ is beyond the accessible time window
($\tau_{\rm M}>10^7$). 

Since as $T\rightarrow 0$ the system enters a glassy state
\cite{kfsPRL2008}, the efficiency $\tau_{\rm M}/
\tau_{\rm W}$ grows at lower $T$ allowing the evaluation of
thermodynamics averages even at $T\ll T_C$ \cite{mazza}.
In particular, the cluster MC algorithm turns out to be very efficient
when approaching the Widom line in the vicinity of the LLCP, with an 
efficiency of the order of $10^4$. 
We plan to analyze in a systematic way how the efficiency $\tau_{\rm M}/
\tau_{\rm W}$ grows on approaching the LLCP.
This result is
well known for the standard liquid-gas critical point \cite{wolff}
and, on the basis of our results, could be extended also to the LLCP.
However, this analysis is very expensive in terms of CPU time and goes
beyond the goal of the present work.
Nevertheless, the percolation analysis, presented in the next section,
helps in understanding the physical reason for this large efficiency. 

The efficiency 
is a consequence of the fact that the average size of Wolff's clusters changes
with $T$ and $P$ in the same way as the average size of the 
regions of correlated molecules \cite{coniglio}, i.e. a Wolff's
cluster statistically represents a region of correlated molecules.
Moreover, the mean cluster size diverges 
at the critical point with 
the same exponent of the Potts magnetic susceptibility
\cite{coniglio}, and the clusters percolate at the critical point, as we will 
discuss in the next section.

\section{Percolating clusters of correlated molecules}

\begin{figure}
\includegraphics[width=0.42\textwidth,height=0.38\textwidth]{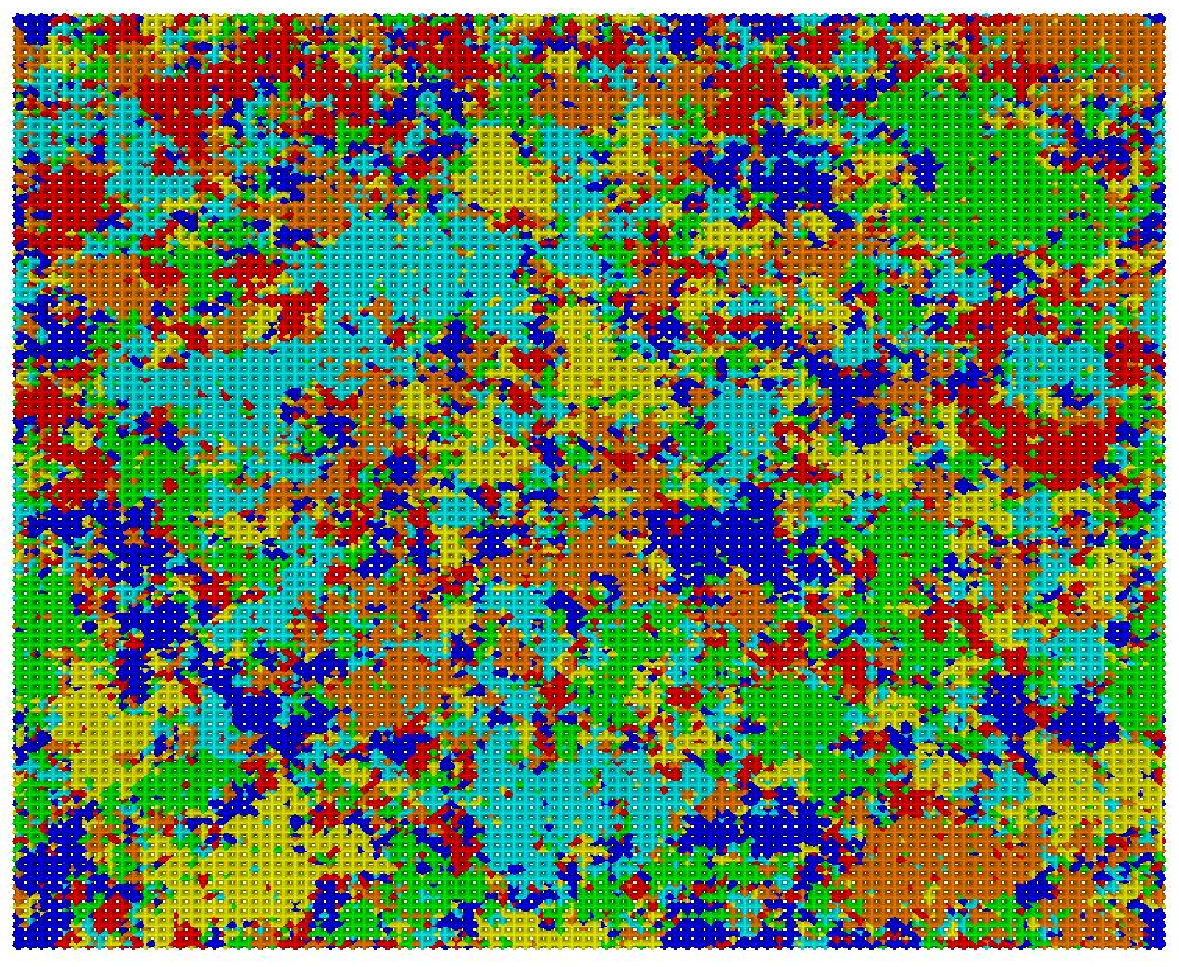}
\includegraphics[width=0.427\textwidth,height=0.38\textwidth]{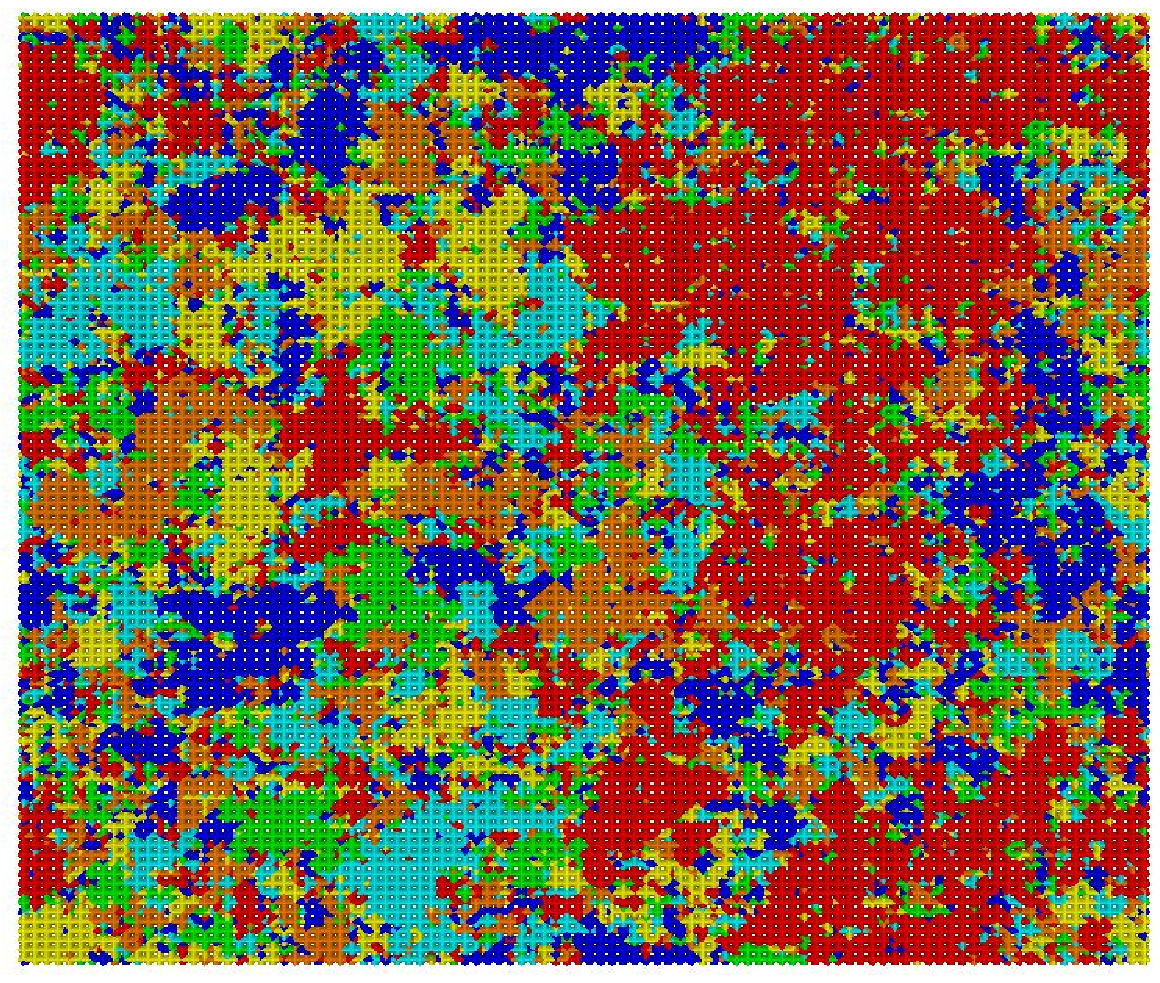}
\includegraphics[width=0.42\textwidth,height=0.38\textwidth]{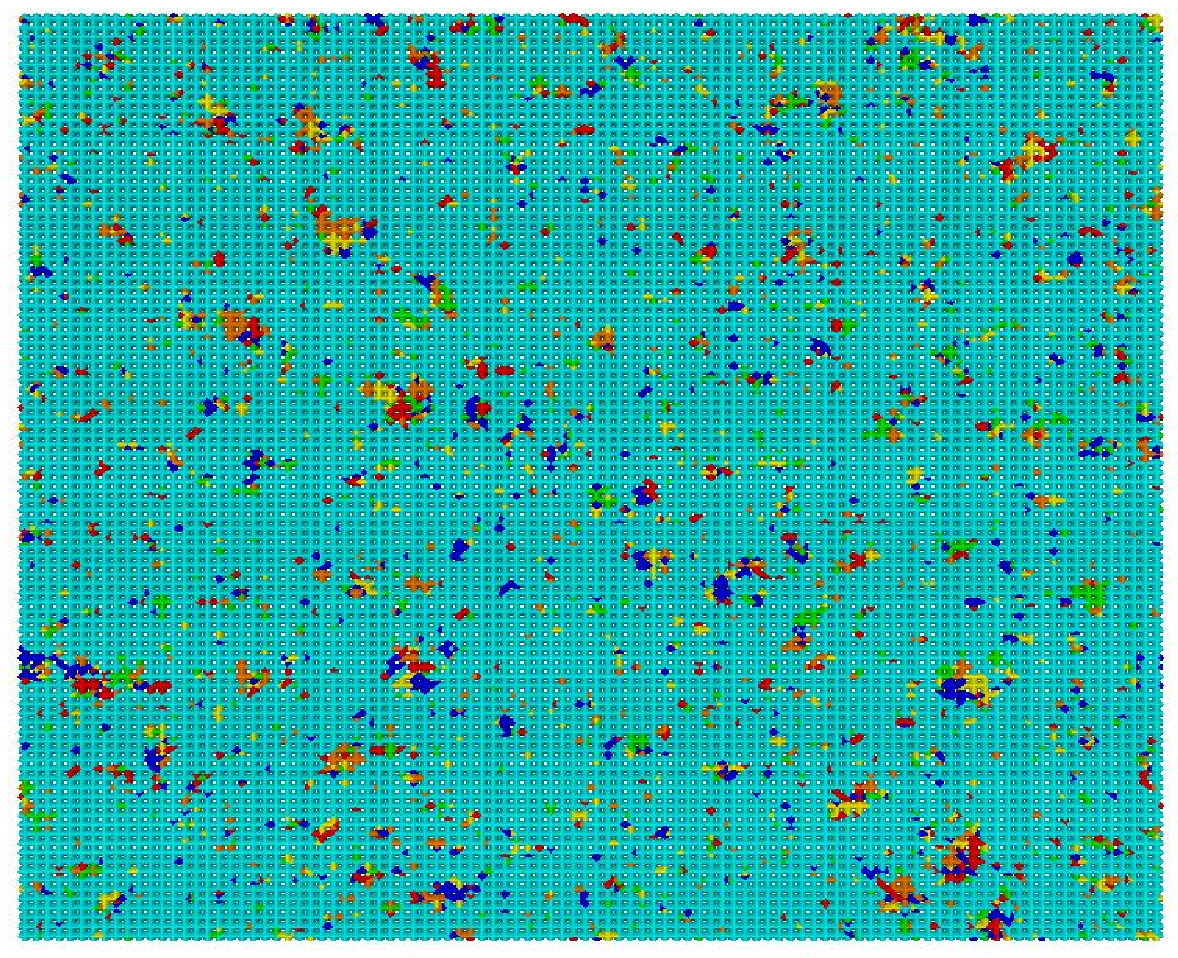}
\caption{Three snapshots of the system with $N=100\times 100$, showing
the Wolff's clusters 
of correlated water molecules. For each molecule we show the states of
the four arms and associate different colors to different arm's states.
The state points are at pressure close to the critical value $P_C$
($Pv_0/\epsilon=0.72\simeq P_Cv_0/\epsilon$) and 
$T>T_C$ (top panel, $k_BT/\epsilon=0.0530$),
$T\simeq T_C$ (middle panel, $k_BT/\epsilon=0.0528$),
$T<T_C$ (bottom panel, $k_BT/\epsilon=0.0520$), showing the onset of
the percolation at $T\simeq T_C$.
\label{clus}}
\end{figure}

The efficiency of the Wolff's cluster algorithm is a consequence of
the exact relation between the average size of the finite clusters
and the average size of the regions of thermodynamically correlated molecules. 
The proof of this relation at any $T$ derives straightforward from the
proof for the case of Potts variables \cite{coniglio}. This relation
allows to identify 
the clusters built during the MC dynamics with the correlated
regions and emphasizes (i) the appearance of heterogeneities in
the structural correlations \cite{hetero}, and (ii) the onset of
percolation of the 
clusters of tetrahedrally ordered molecules at the LLCP
\cite{percolation}, as shown in 
Fig.~\ref{clus}.

A systematic percolation analysis \cite{clusters} is beyond the goal
of this report, however configurations such as those in
Fig.~\ref{clus} allow the following qualitative considerations.
At $T>T_C$ the average cluster size is much smaller than the system
size. Hence, the structural correlations among the molecules extends
only to short distances. This suggests that the correlation time of a
local dynamics, such as Metropolis MC or molecular dynamics, would be
short on average at this temperature and pressure.
Nevertheless, the system appears strongly
heterogeneous with the coexistence of large and small clusters,
suggesting that the distribution of correlation times evaluated among
molecules at a given distance could be strongly heterogeneous.  
The clusters appear mostly compact but with a fractal surface,
suggesting that borders between clusters can rapidly change. 

At $T\simeq T_C$ there is one large cluster, in red on the right of the
middle panel of Fig.~\ref{clus}, with a linear size comparable to the system
linear extension and spanning in the vertical direction. The
appearance of spanning clusters shows the onset of the
percolation geometrical transition. At this state
point the correlation time of local, such as Metropolis MC dynamics
or molecular dynamics 
would be very slow as a consequence of the large extension of the
structurally correlated region. On the other hand, the correlation
time of the Wolff's cluster dynamics is short because it changes in
one single MC step the state of all the molecules in clusters, some of
them with very large size.
Once the spanning cluster is formed,
it breaks the symmetry of the system and a strong effective field acts
on the molecules near its border to 
induce their reorientation toward a tetrahedral configuration with respect
the molecules in the spanning cluster. 

As shown in Fig.3, the spanning cluster appears as
a fractal object, with holes of any size. The same large distribution
of sizes characterizes also the finite clusters in the system. The
absence of a characteristic size for the clusters (or the holes of the
spanning cluster) is the consequence of the fluctuations at any
length-scale, typical of a critical point.

At $T<T_C$ 
the majority of
the molecules belongs to a single percolating cluster that represents
the network of tetrahedrally ordered molecules. 
All the other clusters are small, with a finite size that 
corresponds to the regions of correlated molecules.  
The presence of many
small clusters gives a qualitative idea of the heterogeneity of the
dynamics at these temperatures.

\section{Summary and conclusions}

We describe the numerical solution of mean field equations and the 
implementation of the Wolff's cluster MC
algorithm for a cell model for liquid water. 
The mean field approach allows us to estimate in an approximate way the
phase diagram of the model at any state point predicting intriguing
new results at very low $T$ \cite{mazza}. 

To explore the state points
of interest for these predictions the use of standard simulations,
such as molecular dynamics or Metropolis MC, is not
effective due to the onset of the glassy dynamics \cite{kfsPRL2008}. 
To overcome this problem and access the deeply supercooled region of
liquid water, we adopt the Wolff's cluster MC
algorithm. This method, indeed, allows to greatly accelerate the
autocorrelation time of the system. Direct 
comparison of Wolff's dynamics with 
Metropolis dynamics in the vicinity of the liquid-liquid critical
point shows a reduction of the autocorrelation time of a factor at
least $10^4$. 

Furthermore, the analysis of the clusters generated during the Wolff's
MC dynamics allows to emphasize how the regions of tetrahedrally
ordered molecules build up on approaching the liquid--liquid critical
point, giving rise to the backbone of the tetrahedral hydrogen bond
network at the phase transition \cite{percolation}. The coexistence of
clusters of 
correlated molecules with sizes that change with the state point gives a rationale for the heterogeneous dynamics observed in supercooled
water \cite{hetero}.

\section{Acknowledgments}

We thank Andrew Inglis for introducing one of the authors (MGM) to
VPython, Francesco Mallamace for discussions,
NSF grant CHE0616489 and Spanish MEC grant FIS2007-61433 for support.


\begin{thebibliography}{50}
\bibitem{note}
The model does not differentiate ``donor'' molecule and ``acceptor''
molecule in the hydrogen bond definition. This simplification
increases the number of possible bonded configurations, hence
increases the entropy associated to the local tetrahedral
configurations. A simple modification of the model could explicitly
take into account this feature, however the comparison of the results
from the present version of the model with experiments and 
simulations from more detailed models shows good qualitative agreement.
\bibitem{note2}
The results of \cite{clusters,coniglio} guarantee that the cluster
algorithm described here 
satisfies the detailed balance and is ergodic. Therefore, it is
a valid Monte Carlo dynamics.
\end{thebibliography}


\end{document}